\documentclass[aps,prd,showpacs,amsmath,nofootinbib,
superscriptaddress,twocolumn,preprintnumbers]{revtex4}
\usepackage{natbib}
\usepackage{amsmath}
\usepackage{amssymb}
\usepackage{graphicx}
\usepackage{color}
\usepackage{amsmath}
\usepackage{amsthm}
\usepackage{slashed}
\usepackage{soul}
\begin{document}

\title{Conformal Dilatonic Cosmology}

\author{Meir Shimon}
\affiliation{School of Physics and Astronomy, 
Tel Aviv University, Tel Aviv 69978, Israel}
\email{meirs@wise.tau.ac.il}

\begin{abstract}
Gravitation and the standard model of particle physics are 
incorporated within a single conformal scalar-tensor theory, 
where the scalar field is complex. The Higgs field has a 
dynamical expectation value, as has the Planck mass, but 
the relative strengths of the fundamental interactions are 
unchanged. Initial cosmic singularity and the horizon problem 
are avoided, and spatial flatness is natural. 
There were no primordial phase transitions; consequently, no 
topological defects were produced. Quantum excitations of the 
dilaton phase induced a slightly red-tilted spectrum of gaussian 
and adiabatic scalar perturbations, but no analogous primordial 
gravitational waves were generated. Subsequent cosmological 
epochs through nucleosynthesis are as in standard cosmology. 
A generalized Schwarzschild-de Sitter metric, augmented with 
a linear potential term, describes the exterior of stars and 
galaxies, such that there is no need for dark matter on 
galactic scales.
\end{abstract}

\maketitle

{\bf Introduction}.-
The standard cosmological model has been very successful 
in {\it parametrically} fitting a wide range of cosmological 
observations over a vast dynamical range extending from 
the big bang nucleosynthesis (BBN) to the Hubble scale, e.g. [1, 2]. 
Our understanding of the early universe down to (at least) the BBN era, 
is based on well-understood and experimentally established physics. 

However, current understanding of the {\it very} early universe 
(energies TeV and higher) lacks direct experimental confirmation. 
A major underpinning of standard cosmology is the primordial 
inflationary phase thought to have ended at energies as high 
as $O(10^{16})$ GeV, e.g. [3-6]. A few fine-tuning (naturalness) 
and conceptual problems generically afflict inflation, e.g. [7-10], 
but its exceptional role in explaining and predicting a variety of 
cosmological phenomena is indeed remarkable. 

On the largest scales, and rather recently, a non-clustering mysterious 
vacuum-like species, which is many orders of magnitude 
smaller than expected on theoretical grounds from a vacuum energy, e.g. [11], 
has come to dominate the background cosmological 
dynamics, e.g. [12, 13]. Additionally, 
on cosmological down to sub-galactic scales cold dark matter (CDM) is required for 
an observationally viable cosmological model. In the absence of this species 
the global spatial flatness, non-Keplerian galactic rotation curves, 
strong gravitational lensing, 
and the observed abundance of nonlinear structure, could not be explained 
by a (largely) baryon-dominated cosmological model.

Moreover, in spite of the success of the standard cosmological model the 
microphysics of most of the cosmic energy budget is either unknown or 
fine-tuned to explain observations. Judging a physical theory based only on its 
(indeed remarkably small) number of free parameters is arguably unsatisfactory.

In this {\it letter} we explore a few cosmological implications of promoting fundamental 
physical constants, e.g. particle masses, Newton gravitational constant $G$, etc., to scalar 
fields. We adopt a conformal dilatonic framework for describing 
the fundamental interactions and formulate cosmology in a (field) frame-independent fashion. 
Important conceptual differences between our approach and standard cosmology, and 
departures in the physics of the early universe and galactic dynamics, are briefly highlighted.
For a more complete and comprehensive presentation of the new approach 
summarized here, see [14]. Throughout, we adopt a mostly positive metric signature. 

{\bf Conformal Gravity and the Standard Model}.-
Focusing on the gravitational sector 
of the fundamental interactions first, we consider a (complex) 
scalar-tensor theory of gravity that is formulated in 
terms of the action
\begin{eqnarray}
\mathcal{I}_{gr}=\int\left[\frac{1}{6}|\phi|^{2}R
+\phi_{,\mu}\phi^{*,\mu}+\mathcal{L}_{M}(|\phi|)\right]\times\sqrt{-g}d^{4}x,
\end{eqnarray}
merely a generalization of the action discussed 
in [15] to the complex field case,
where $g_{\mu\nu}$, $R$ and $\mathcal{L}_{M}$ are the metric, 
curvature scalar, and matter lagrangian density, respectively. 
The dilaton field $\phi$ has a non-positive kinetic term, 
a feature which is usually considered a problem, but here we show that 
only its phase perturbation is a genuine degree of freedom subject to quantization in 
conformal dilatonic gravity.  
The matter lagrangian $\mathcal{L}_{M}(|\phi|)$ may also contain terms $\propto|\phi|^{4}$ 
which function as vacuum-like energy contributions (thereby effectively promoting the 
cosmological constant to a dynamical quantity).
Eq. (1) is invariant under the generalized conformal (Weyl) transformation 
$\phi\rightarrow\phi/\Omega$, $g_{\mu\nu}\rightarrow\Omega^{2}g_{\mu\nu}$, 
and $\mathcal{L}_{M}\rightarrow \mathcal{L}_{M}/\Omega^{4}$ 
where $\Omega(x)$ is an arbitrary function of spacetime, e.g. [16]. 
With $\mathcal{L}_{M}$ independent of $\phi$ this theory is a version of a 
particular Brans-Dicke (BD) theory, with the dimensionless BD parameter 
$\omega_{BD}=-3/2$ [15]. In the presence of 
$\mathcal{L}_{M}(|\phi|)$ it falls in the category 
of Bergmann-Wagoner scalar-tensor theories [17, 18].

Variation of Eq. (1) with respect to $g_{\mu\nu}$ and $\phi$ results in 
generalized Einstein equations, as well as equations for the scalar fields [19]. 
Combining these equations results in a generalized local energy momentum (non-) 
conservation law
\begin{eqnarray}
T_{M,\mu;\nu}^{\nu}&=&\mathcal{L}_{M,\phi}\phi_{,\mu},
\end{eqnarray}
where $T_{M,\mu\nu}\equiv
\frac{-2}{\sqrt{-g}}\frac{\delta(\sqrt{-g}\mathcal{L}_{M})}{\delta g^{\mu\nu}}$ 
is the matter energy-momentum tensor.
Indeed, energy-momentum is clearly not conserved when Newton constant $G$, 
the cosmological constant $\Lambda$, or particle masses, 
are promoted to spacetime-dependent fields. 
Massless particles still travel along geodesics in this theory but massive 
particles do not. This fact is responsible for, e.g., cosmological redshift in a 
comoving frame where the metric field is static, and non-Keplerian behavior 
of galactic rotation curves -- phenomena that are usually attributed 
to space expansion and galactic CDM, respectively [14]. 

Embedding the standard model (SM) of particle physics in this 
conformal theory of gravity is straightforward, and is summarized by 
the following action that accounts for both 
the gravitational and the SM interactions 
\begin{eqnarray}
\mathcal{I}_{tot}&=&\int d^{4}x\sqrt{-g}
\left[\frac{1}{6}(|\phi|^{2}-H^{\dagger}H)R\right.\nonumber\\
&+&\left.(\phi^{*\mu}\phi_{\mu}-D^{\mu}H^{\dagger}D_{\mu}H)
-\lambda_{SM}(H^{\dagger}H-v^{2})^{2}\right.\nonumber\\
&+&\left.\mathcal{L}_{SM}(v, \boldsymbol{\psi}, {\bf A}_{\mu}, g^{\mu\nu})\right].
\end{eqnarray}
Here, $H$ is the Higgs isospin doublet, 
$\lambda_{SM}$ is its dimensionless self-interaction constant, and $v\equiv\alpha|\phi|$ 
is its vacuum expectation value (VEV) where $\alpha=O(10^{-16})$ is the 
dimensionless ratio of the Higgs VEV and the dynamical 
(reduced) Planck mass. 
The lagrangian $\mathcal{L}_{SM}$ is a function of $v$, fundamental 
fermions $\boldsymbol{\psi}$, gauge bosons ${\bf A}_{\mu}$, and 
the metric $g_{\mu\nu}$. Eq. (3) differs from a similar action 
discussed in [22] by one crucial aspect; here, unlike in [22], $\phi$ 
is complex. In addition, unlike in [22] and similar works, we never 
fix $\phi$ to a constant value; masses (essentially $v$), $G$, 
and $\Lambda$ are dynamical. 

{\bf Cosmological Model}.-
Defining $\phi\equiv\rho e^{i\theta}$, $\tilde{a}\equiv a\rho$ and 
$\tilde{\mathcal{H}}\equiv\tilde{a}'/\tilde{a}$ 
with $f'\equiv\frac{df}{d\eta}$ for any $f$, where 
$\eta$ is the conformal time related to cosmic time $t$ via $dt\equiv a(\eta)d\eta$, 
and $a(\eta)$ is the scale factor describing the time-dependence of the 
Friedmann-Robertson-Walker (FRW) metric 
$g_{\mu\nu}=a^{2}\cdot diag(-1, \frac{1}{1-Kr^{2}}, r^{2}, r^{2}\sin^{2}\theta)$ 
in spherical coordinates and conformal time units, 
the field equations derived from Eq. (1) can be cast in a manifestly 
frame-invariant fashion [14]
\begin{eqnarray}
\tilde{\mathcal{H}}^{2}+K&=&\tilde{a}^{2}\tilde{\rho}_{M}
+\lambda\tilde{a}^{2}-\theta'^{2}\\
\tilde{\mathcal{H}}'+\tilde{\mathcal{H}}^{2}+K&=&
\frac{(1-3w_{M})}{2}\tilde{a}^{2}\tilde{\rho}_{M}
+2\lambda\tilde{a}^{2}+\theta'^{2}.
\end{eqnarray}
Here, $K$ is the spatial curvature, 
the matter energy-momentum tensor of a perfect fluid is 
$T_{M,\mu}^{\nu}=\rho_{M}\cdot diag(-1,w_{M},w_{M},w_{M})$, 
where $w_{M}\equiv\rho_{M}/P_{M}$ is the equation of state (EOS), 
and $\rho_{M}$ \& $P_{M}$ are the energy density and pressure, respectively.
The analog of the energy density scales as 
$\tilde{\rho}_{M}\equiv\sum_{i}\tilde{\rho}_{M_{0,i}}\tilde{a}^{-3(1+w_{M,i})}$, 
where the index $i$ runs over the species.
The effective analog energy densities associated with $\theta$ and $K$ 
are $\tilde{\rho}_{\theta}\equiv -\theta'^{2}/\tilde{a}^{2}$ and 
$\tilde{\rho}_{K}\equiv -K/\tilde{a}^{2}$, respectively.
In addition to matter we include a cosmic term with an effective $w_{\lambda}=-1$ 
which is characterized by 
$\tilde{T}_{\lambda,\mu}^{\nu}=\lambda\cdot diag(-1,1,1,1)$, 
where $\lambda$ is a fixed dimensionless constant. 
The $\lambda\tilde{a}^{2}$ term appearing in Eq. (4) drives 
a very early conformal phase of evolution and is also responsible for 
the recent vacuum-like dominated era.

The generalized Friedmann equations, Eqs. (4) \& (5), are invariant under any simultaneous 
change of $\rho$ and $a$ (i.e. particle masses and the scale factor, respectively) that 
leaves their product, $\tilde{a}$, unchanged. Note that the EOS 
of perfect fluids does not change under conformal transformations. 
Since the global $U(1)$ symmetry of Eq. (1) implies that $\theta'\propto\tilde{a}^{-2}$, the 
dilatonic phase term $\propto\theta'^{2}$ in Eqs. (4) \& (5) can be described as an 
effective stiff matter ($w_{\theta}=1$) carrying negative energy density.
In [14] it is shown that the linear order perturbation equations 
governing the metric and matter perturbations are frame-invariant as well. 
Moreover, consistency of the perturbation equations associated with 
the dilaton with the perturbed trace of the Einstein equations, implies 
that (cosmological) metric perturbations in 
the theory described by Eq. (1) must be adiabatic, 
i.e. $\delta P_{M}=w_{M}\delta\rho_{M}$ [14].
In addition, perturbations of the modulus $\delta\rho$ can be systematically absorbed 
in `renormalized' metric and matter perturbations. In particular, this implies that 
real fields, such as the inflaton, cannot seed metric perturbations if conformally coupled 
to gravity. For this reason, quantum and thermal fluctuations of $\rho$ 
are absent, implying no catastrophic particle production is associated with the negative 
kinetic term of Eq. (1) and no primordial phase transitions occurred, respectively.
The observed non-clustering of the DE contribution 
to the cosmic energy budget is consistent with $\tilde{\rho}_{\lambda}=\lambda$ 
being a dimensionless constant, not a field. In any case, the DE terms appearing 
in Eqs. (4) \& (5) are $\propto\lambda\tilde{a}^{2}$ and since $\tilde{a}$ is 
unperturbed then the effective DE contribution to the perturbed Friedmann equation 
identically vanishes.
As shown below, quantum fluctuations of the phase $\theta$ that seed scalar perturbations 
are governed by a Mukhanov-Sasaki-like wave equation.

In contrast to the standard cosmological model where cosmic time, $t$, is effectively 
replaced by $\eta$ for massless particles, in our proposed cosmological model the `cosmic clock' 
ticks universally for both massless and massive particles. Time is 
parametrized by $\eta$, with cosmological redshift generally explained by a combination 
of space expansion and dynamical masses, i.e. varying Rydberg `constant'. 
Redshift is explained solely by the latter effect in the comoving frame.

The very early universe scenario described in this {\it letter} 
begins with a very large {\it deflating} $\tilde{a}$ in a `conformal era' 
when the right hand sides of Eqs. (4) \& (5) are dominated by the $\propto\lambda\tilde{a}^{2}$ 
terms. Therefore, in the absence of dimensional constants 
$\tilde{a}\propto(\eta-\eta')^{-1}$, where $\eta'$ is an arbitrarily 
negative integration constant (since our cosmological model is 
non-singular as shown below), scales according to its canonical dimension, 
$[\tilde{a}]=[\rho]=length^{-1}$, and the spontaneously broken 
conformal symmetry is `restored'.  

An important aspect of the conformal era ($w_{M}=-1$) 
in the cosmological context, and within a theory featuring 
a conformally-coupled complex scalar field,  
is the generation of gaussian and adiabatic 
scalar metric perturbations which are characterized 
by a flat spectrum [23]. The mechanism proposed here is different in a few crucial 
aspects, e.g. unlike in [23] we consider the effect of $\theta'\neq 0$ 
as well as coupling of phase- to metric-perturbations. These differences 
result in fundamentally different explanations for the tilt of the 
power spectrum and the adiabacity of scalar perturbations [14], and avoids 
a few difficulties with [23] stemming from coupling 
of $\delta\rho$ to $\delta\theta$.
Working in the shear-free gauge, and neglecting stress anisotropy, 
the linear perturbation equations in the very early conformal epoch 
satisfy [14]
\begin{eqnarray}
&&\varphi''+6\tilde{\mathcal{H}}\varphi'+(q^{2}+6\lambda\tilde{a}^{2})\varphi=0,\\
&&\delta\theta=(\varphi'+\tilde{\mathcal{H}}\varphi)/(3\theta'),
\end{eqnarray}
where $\varphi$ is the `renormalized' Newtonian potential, 
$q^{2}\equiv k^{2}-8K$.  
The phase dynamics ($\theta'\neq 0$) plays a crucial 
role in mediating phase perturbations 
$\delta\theta$ to scalar metric perturbations (Eq. 7).
Neglecting spatial curvature, and using $\tilde{a}=(\sqrt{\lambda}\eta)^{-1}$ 
during the conformal era (where we set an integration constant to 0 for simplicity), 
the general solution of Eq. (6) is 
$\varphi=c_{1}\eta^{\frac{7}{2}}J_{5/2}(k\eta)+c_{2}\eta^{\frac{7}{2}}Y_{5/2}(k\eta)$, 
where $J_{5/2}(k\eta)$ and $Y_{5/2}(k\eta)$ are the Bessel functions of the 
first and second kinds, respectively.
Requiring the appropriate Bunch-Davis vacuum for $\delta\theta$ at $k\eta\gg 1$ 
determines the coefficients $c_{1}$ and $c_{2}$ and results 
in $\varphi\propto k^{-1}\eta^{5/2}H_{5/2}^{(1)}(k\eta)$ (where $H_{5/2}^{(1)}$ is the Hankel 
function of the first kind and of order 5/2), which in the $k\eta\ll 1$ 
limit corresponds to $k^{3}P_{\varphi}(k)=k^{-4}$. This translates 
to $k^{3}P_{\delta{\rho_{M}}}(k)=constant$ by virtue of the Poisson equation.
Here $P_{\varphi}$ \& $P_{\delta{\rho_{M}}}$ are the power spectra of metric and density 
perturbations, respectively.
A pseudo-conformal epoch with $w_{M}\gtrsim -1$ 
similarly results in regular perturbation modes at the bounce 
which are characterized by a red-tilted power spectrum. 
In contrast, $w_{M}\lesssim -1$ corresponds to 
singular perturbation modes at the bounce, thereby 
undermining the underlying homogeneity of the cosmological model
(and is therefore not a viable solution), 
and are described by a blue-tilted spectrum [14]. 

This mechanism of phase-perturbation-induced metric perturbations 
is exclusive to perturbations of the scalar type. A positive 
detection of primordial gravitational waves (PGW) would definitely 
challenge the proposed mechanism. Since Eq. (6) is linear in 
$\varphi$ there is no mode-mode coupling and because we 
assume (as is common in inflationary models) that the underlying 
fluctuating $\theta$ is a quantum vacuum, and is therefore gaussian, 
then one expects the induced metric perturbations to be 
likewise gaussian. 

Curvature and the effective stiff matter contributions may still be 
dynamically important at sufficiently small $\tilde{a}$. 
Assuming all matter was relativistic (near the minimal $\tilde{a}$) 
with energy density $\rho_{r}$, and integrating Eq. (4) (while ignoring 
the potential term) then results in 
\begin{eqnarray}
\tilde{a}^{2}&\approx&\left(\frac{\rho_{\theta,*}}{K}
+\left(\frac{\rho_{r,*}}{2K}\right)^{2}\right)^{1/2}
\cosh(2\sqrt{-K}(\eta-\eta_{*}))\nonumber\\
&-&\frac{\rho_{r,*}}{2K}
\end{eqnarray}
and its minimum is attained at $\eta=\eta_{*}$, where 
(recall that) $\rho_{\theta}\leq 0$ and $K<0$. 
When the $\rho_{\theta}$ and the effective $\rho_{K}$ terms become subdominant 
to $\rho_{r}$, BBN begins, followed by all standard cosmological epochs, 
including radiation-matter 
equality, recombination, and the recent acceleration. 
In case that $K=0$ Eq. (4) integrates to 
$\tilde{a}^{2}=\rho_{r,*}(\eta-\eta_{*})^{2}-\rho_{\theta,*}/\rho_{r,*}$. 
In could be easily shown that in this case not only $\eta$ is extended 
to $-\infty$, but also $t=\int a(\eta)d\eta$ is. In other words, both 
timelike and null geodesics are freely extended through the non-singular 
bounce. Since the initial cosmic singularity is avoided no matter generation 
mechanism is required -- spacetime and matter always existed.
In addition, it is straightforward to show that $\rho_{\theta,*}$ 
can be naturally chosen such that scalar metric perturbations are finite 
at the bounce and its vicinity [14], thereby not undermining the underlying 
homogeneity on cosmological scales. By selecting $\rho_{\theta,*}$ sufficiently 
small the post-bounce era starts at sufficiently large number densities to guarantee very effective
double-Compton and bremsstrahlung processes at its minimum $\tilde{a}$ value and thereby the 
effective thermalization of the CMB is guaranteed. Structure formation history is exactly 
as in standard cosmology since both the background 
equations (Eqs. 4 \& 5) and the perturbation 
equations [14] are unchanged in the relevant post radiation-dominated era.

The flatness problem, which is addressed in standard cosmology by inflation, does not exist in 
our early universe scenario. The problem essentially arises in the standard model 
since space monotonically expands, but in a non-singular bouncing scenario such as 
the one advocated here no fine-tuning of the curvature is required. 
In the deflationary phase the matter 
density grows in proportion to the energy density associated with curvature. Since 
the model is symmetric in $\tilde{a}$ this implies that for curvature to dominate over 
matter at present the matter density should have been extremely fine-tuned to zero 
at $\eta=-\infty$. From that perspective, matter domination at any finite time 
is actually an attractor rather than an unstable point.

On galactic and sub-galactic 
scales (and possibly on extra solar system scales) 
we consider a spherically symmetric static line element (with conformal time)
\begin{eqnarray}
ds^{2}=-B(r)d\eta^{2}+B^{-1}(r)dr^{2}+r^{2}(d\theta^{2}+\sin^{2}\theta d\varphi^{2}).
\end{eqnarray}
The metric $B(r)$ is 
determined (up to conformal transformations of $g_{\mu\nu}$ and $\rho$) 
by the field equations, in vacuum and with no cosmological constant, to be [14]
\begin{eqnarray}
\rho&=&\frac{\rho_{0}}{1+\gamma r/(2-3\beta\gamma)}\\
B&=&(1-3\beta\gamma)-\frac{\beta(2-3\beta\gamma)}{r}+\gamma r-\kappa r^{2} ,
\end{eqnarray}
where $\rho_{0}$, $\beta$ and $\gamma$ are integration constants. 
We note that $B(r)$ coincides with the 
corresponding quantity obtained in fourth order Weyl gravity [20, 21]. 
In accord with what has been done in [20, 21], we considered the limit $\beta\gamma\ll 1$. 
Note that $\kappa$ plays the role of an effective cosmological constant 
($\Lambda=3\kappa$) even though Eqs. (10) \& (11) are obtained as 
a {\it vacuum solution} of the field equations; this can be understood if 
the cosmological constant is viewed (in static spacetimes) 
as an arbitrary integration constant.

As noted in, e.g., [20, 21] the linear term appearing in Eq. (11) may account for the 
shapes of galactic rotation  
curves and strong lensing data with no recourse to CDM on these scales. This would 
set a lower bound on the CDM particle mass of $\gtrsim 10^{-22} eV/c^{2}$ [14] if it 
is assumed that CDM does cluster on galaxy cluster scales, as may be implied by bullet-like 
clusters. We note 
that the metric and scalar field in Eqs. (10) \& (11) are determined only up to a conformal 
rescaling which we determine to be 
$g_{\mu\nu}\rightarrow\tilde{a}^{2}g_{\mu\nu}$ and $\rho\rightarrow\rho/\tilde{a}$, 
commensurate with observations of emission by sources 
residing within gravitationally bound objects [14].

{\bf Summary}.-
This work advocates abandoning the standard units convention that underlies 
general relativity (GR) and the SM 
of particle physics -- local energy-momentum conservation -- in favor 
of local scale invariance, i.e. conformal (Weyl) symmetry. 
One might argue that forcing energy-momentum 
conservation on galactic scales required cosmologists to introduce CDM in order to explain 
the observed anomalous rotation curves and strong lensing data. While 
invoking the CDM hypothesis has been rather successful in 
{\it parametrically} fitting observations to GR predictions, the essence of CDM remains elusive. 

Whereas conformal dilatonic gravity naturally accommodates 
quartic potential inflation, perturbations of scalar fields moduli are 
systematically absorbed in renormalized metric and matter perturbations. 
Therefore, they cannot be used to explain the seed 
density perturbations if the inflaton field is conformally coupled to gravity. 
We consider an alternative bouncing cosmological 
scenario which is symmetric in $\tilde{a}$ around the bounce. 
The deflating pre-bounce era involves a conformal cosmic 
epoch followed (possibly) by a (negative) curvature-dominated (CD) era, 
nonrelatvistic matter, radiation, and an effective `stiff' matter component; 
the latter reflects the dynamics of the transversal mode of the complex 
scalar field, and is characterized by an effective negative energy density, 
thereby providing a `centrifugal barrier' that is responsible for the bounce 
at a finite $\tilde{a}$.

The deflating pseudo-conformal 
evolutionary phase ($w_{M}\gtrsim -1$), $\tilde{a}\propto\eta^{-1}$, 
where $\tilde{a}$ rolls down its (nearly) 
quartic potential dominates the evolution for sufficiently 
large $\tilde{a}$. Perturbations of the dilaton phase induce a 
nearly flat, red-tilted spectrum of gaussian and adiabatic scalar 
metric perturbations. 
No analogous production of either PGW or vector perturbations is expected, rendering this 
mechanism (and possibly the entire framework) falsifiable. 
The flatness problem is naturally addressed by the 
non-singular {\it bouncing} scenario; matter domination over curvature is 
an attractor point at any finite time. In addition, the horizon problem is avoided by the fact 
that the model is non-singular. No primordial phase transitions occur 
in conformal dilatonic gravity, and consequently no primordial relic problem arises 
in the first place. These cosmic epochs are subsequently followed by the conventional 
radiation- and matter-domination, recombination, etc. Structure 
formation history is unchanged. 

Conformal dilatonic gravity admits spherically symmetric vacuum solutions for 
a modified Schwarzschild-de Sitter spacetime augmented by a linear potential term. 
When applied to galactic scales, this approach results in significant 
departures from standard interpretations of observations. 
This pertains, in particular, to our understanding of the nature of cosmological redshift, 
CDM, and DE. The implication is that CDM may not be required on galactic 
and sub-galactic scales, but may be required on galaxy cluster 
scales and larger for a proper phenomenological description of `bullet'-like systems. 
This fact alone 
already sets a lower bound $m_{CDM}\gtrsim 10^{-22} eV/c^{2}$ on the mass of CDM particles.

We have shown that the dynamics of conformally-coupled complex scalar fields 
can account for cosmological redshift in a (field) frame-independent fashion, explain 
away the horizon and flatness problems, and avoid initial cosmic singularity and primordial 
relics. Additionally, our theoretical formulation 
naturally explains the spectrum of primordial density perturbations, 
their gaussianity and adiabacity, 
removes the need for invoking CDM on galactic scales, provides 
the dynamic VEV for the Higgs field (up to a large hierarchy constant), and 
thereby all fundamental length (and mass) scales, all in a single unified framework that 
underscores the unique role played by conformal symmetry, possibly an overarching 
symmetry of the four fundamental interactions.

{\bf Acknowledgments}.-
The author is indebted to Yoel Rephaeli for numerous constructive, critical, and 
thought-provoking discussions which were invaluable for this work.

\end{document}